\documentclass[conference]{IEEEtran}
\ifCLASSINFOpdf
 \usepackage[pdftex]{graphicx}
\DeclareGraphicsExtensions{.pdf}
\else
\fi
\begin{document}
%
\title{The role of distances in the World Trade Web}



\author{\IEEEauthorblockN{Francesco Picciolo, Franco Ruzzenenti, Riccardo Basosi}
\IEEEauthorblockA{Department of Chemistry\\
University of di Siena\\
Siena, Via Aldo Moro 1, 53100 Siena (Italy)\\
Email: picciolo@unisi.it, ruzzenenti@unisi.it,\\ 
basosi@unisi.it}
\and
\IEEEauthorblockN{Tiziano Squartini, Diego Garlaschelli}
\IEEEauthorblockA{Instituut-Lorentz for Theoretical Physics\\
Leiden Institute of Physics, University of Leiden\\
Niels Bohrweg 2, 2333 CA Leiden (The Netherlands)\\
Email: squartini@lorentz.leidenuniv.nl,\\
garlaschelli@lorentz.leidenuniv.nl}
}


\maketitle

\begin{abstract}
In the economic literature, geographic distances are considered fundamental factors to be included in any theoretical model whose aim is the quantification of the trade between countries. Quantitatively, distances enter into the so-called gravity models that successfully predict the weight of non-zero trade flows. However, it has been recently shown that gravity models fail to reproduce the binary topology of the World Trade Web. In this paper a different approach is presented: the formalism of exponential random graphs is used and the distances are treated as constraints, to be imposed on a previously chosen ensemble of graphs. Then, the information encoded in the geographical distances is used to explain the binary structure of the World Trade Web, by testing it on the degree-degree correlations and the reciprocity structure. This leads to the definition of a novel null model that combines spatial and non-spatial effects. The effectiveness of spatial constraints is compared to that of non-spatial ones by means of the Akaike Information Criterion and the Bayesian Information Criterion. Even if it is commonly believed that the World Trade Web is strongly dependent on the distances, what emerges from our analysis is that distances do not play a crucial role in shaping the World Trade Web binary structure and that the information encoded into the reciprocity is far more useful in explaining the observed patterns.
\end{abstract}



%
\IEEEpeerreviewmaketitle

\section{Introduction}

We usually consider networks only from the topological point of view, with the adiacency matrix encoding all the necessary information about the connections between nodes. However, many networks are also embedded into a metric space and vertices have positions described by metric coordinates. In these networks, distances are naturally induced between nodes and geometric proximity represent a novel kind of connectedness to be defined for the vertices. An interesting goal becomes quantifying the influence that geometric distances have on the purely topological connections. A clear example is provided by \emph{geographic distances}.

The role of distances in shaping the World Trade Web (WTW), i.e. the network of import-export trade relationships among
all world countries, enters, in the economic literature, only into the definition of the class of models called \emph{gravity models} \cite{IEEEhowto:giorgio,IEEEhowto:gravity}. The latter, mimicking the equation of the gravitational potential, predict an intensity of trade between two countries, $i$ and $j$, which (in the simplest case) is directly proportional to their Gross Domestic Products (GDPs) and inversely proportional to their geographical distance. So the fundamental ingredients in the economic recipe are the GDPs of the involved countries and the geographic distances between them, disfavouring distant countries to establish intense trading relationships.

Even if gravity models have been proved to be valid to predict the weighted structure of the WTW \cite{IEEEhowto:giorgio2}, three limitations of this approach consist 1) in the exclusively weighted nature of the predicted network, 2) in the trivial topological structure it induces, 3) in the lack of a reciprocity structure of trade-flows between the same nodes that cannot be predictable. In fact, gravity models cannot predict zero trade between any two countries (exactly as the gravity force between any two bodies cannot be zero), thus creating a trivially, fully connected World Trade Web. Moreover, by using only the aforementioned quantities, even if asymmetric flows can be induced (by means of additional parameters: usually exponents), a ``reciprocal flow'' cannot be defined between countries, thus failing to reproduce the strong observed reciprocity \cite{IEEEhowto:mywegrec,IEEEhowto:mygrandcanonical,IEEEhowto:myreciprocity,IEEEhowto:mymotifs} of trade-exchanges. Variations of the gravity models have been defined so far (the so-called \emph{zero-inflated gravity models} \cite{IEEEhowto:giorgio2}) to overcome the first two problems and to be able also to predict the \emph{existence} of a link (and not only its weight, once its existence has been observed). However, the prediction thus obtained does not seem to be good at all \cite{IEEEhowto:giorgio2}, with the consequence that all the topological structure has be known in advance, to succesfully reproduce the observed weights.

In this paper we overcome these limitations, by using a different approach: the exponential random graph formalism. In this theoretical framework, geographic distances are considered as given, exactly as in the previous case, but are used to calculate the probability according to which any two nodes establish a directed connection (a trading relationship, seen as export or import): the only additional information comes from some kind of chosen topological constraint. Moreover, this framework allows us to compare the effectiveness of the distances in explaining the observed patterns with respect to other well known quantities, as the degree sequence and the reciprocity: what turns out is that distances do not add significantly more information to what already predicted by the degree sequence alone which, in turn, is known to be related to the GDPs of the world countries \cite{IEEEhowto:myinterplay,IEEEhowto:mylikelihood}.

The results presented in what follows are about the WTW as considered in its binary, directed representation (BDN, in what follows) as obtained by the database in \cite{IEEEhowto:gledi}. Following \cite{IEEEhowto:myfilling}, our aim is to disentangle spatial and non-spatial effects in the real WTW. To this end, our approach is the comparison of the observed WTW with the prediction of various null models that control either for purely topological effects or for a combination of topological and spatial effects. Finally we introduce a way to quantitatively assess the significance of the information gained by adding geographic distances to non-spatial models of trade.

\section{Null models}

The method we use to introduce null models of the World Trade Web implements a recently proposed procedure \cite{IEEEhowto:mylikelihood,IEEEhowto:mymethod}, developed inside the exponential random graph theoretical famework \cite{IEEEhowto:HL,IEEEhowto:WF,IEEEhowto:newman_expo}. The method is composed by two main steps: the first one is the maximization of the Shannon entropy over a previously chosen set of graphs, $\mathcal{G}$

\begin{equation}
S=-\sum_{G\in\mathcal{G}}P(G)\ln P(G)
\end{equation}

\noindent under a number of imposed constraints \cite{IEEEhowto:mymethod,IEEEhowto:shannon,IEEEhowto:jaynes}, generically indicated as

\begin{equation}
\sum_{G\in\mathcal{G}}P(G)=1,\:\sum_{G\in\mathcal{G}}P(G)\pi_{a}(G)=\langle\pi_a\rangle,\:\forall\:a
\end{equation}

\noindent (note the generality of the formalism, above: $G$ can be a directed, undirected, binary or weighted network). We can immediately choose the set $\mathcal{G}$ as the grandcanonical ensemble of BDNs, i.e. the collection of networks with the same number of nodes of the observed one (say $N$) and a number of links, $L$, varying from zero to the maximum (i.e. $N(N-1)$). This prescription leads to the exponential distribution over the previously chosen ensemble

\begin{equation}
P(G|\vec{\theta})=\frac{e^{-H(G,\:\vec{\theta})}}{Z(\vec{\theta})},
\end{equation}

\noindent whose coefficients are functions of the Hamiltonian, $H(G,\:\vec{\theta})=\sum_{a}\theta_{a}\pi_{a}(G)$, which is the linear combination of the chosen constraints. The normalization constant, $Z(\vec{\theta})\equiv\sum_{G\in\mathcal{G}}e^{-H(G,\:\vec{\theta})}$, is the partition function \cite{IEEEhowto:mymethod,IEEEhowto:newman_expo,IEEEhowto:jaynes}.

The second step prescribes how to numerically evaluate the unknown Lagrange multipliers. Given a real network, $G^*$, Let us consider the log-likelihood function $\ln\mathcal{L}(\vec{\theta})=\ln P(G^*|\vec{\theta})$ and maximize it with respect to the unknown parameters \cite{IEEEhowto:mylikelihood,IEEEhowto:mymethod}. In other words, we have to find the value $\vec{\theta}^*$ of the multipliers satisfying the system

\begin{equation}
\frac{\partial\ln\mathcal{L}(\vec{\theta})}{\partial\theta_a}\bigg|_{\vec{\theta}^*}\equiv0,\:\forall\:a,
\end{equation}

\noindent or, that is the same,

\begin{equation}
\pi_{a}(G^*)=\langle\pi_a\rangle(\vec{\theta}^*)\equiv\langle\pi_a\rangle^*,\:\forall\:a
\label{exp}
\end{equation}

\noindent i.e. a list of equations imposing the value of the expected parameters to be equal to the observed one \cite{IEEEhowto:mylikelihood,IEEEhowto:mymethod}. Note that the term ``expected'', here, refers to the weighted average taken on the grandcanonical ensemble, the weights being the probability coefficients defined above.

So, once the unknown parameters have been found, it is possibile to evaluate the expected value of any other topological quantity of interest, $X$: 

\begin{equation}
\langle X\rangle^*=\sum_{G\in\mathcal{G}}X(G)P(G|\vec{\theta}^*).
\end{equation} 

Because of the difficulty to analytically calculate the expected value of the quantities commonly used in complex networks theory, it is often necessary to rest upon the linear approximation method: $\langle X\rangle^*\simeq X(\langle G\rangle^*)$.

This is a very general prescription, valid for binary, weighted, undirected or directed networks: since the WTW has been considered in its binary, directed representation, the generic adjacency matrix $G$ will be indicated, from now on, with the usual letter $A$; so, $\langle A\rangle^*$ indicates the expected adjacency matrix, whose elements are $\langle a_{ij}\rangle^*\equiv p_{ij}^*$.

The next four subsections will be devoted to the explanation of the null models considered in the present analysis.

\subsection{Directed Configuration Model (DCM)}

The DCM is one of the most used null models in the complex networks literature \cite{IEEEhowto:newman_expo}. The reason why we choose it as our baseline model is that the DCM has been shown to reproduce remarkably well several properties of the WTW, including degree correlations and clustering coefficients \cite{IEEEhowto:pre1}. The DCM Hamiltonian is the following:

\begin{equation}
H(A,\:\vec{\theta})=\sum_{i=1}^N(\alpha_i k_i^{out}+\beta_i k_i^{in})=\sum_{i=1}^N\sum_{j(\neq i)=1}^N(\alpha_i+\beta_j)a_{ij};
\end{equation} 

\noindent the linearity of the constraints (the in-degree and out-degree sequences) in the adjacency matrix elements implies that the probability coefficient for the generic network, $A$, factorizes as a product over the directed pairs of nodes

\begin{equation}
P(A|\vec{\theta})=\prod_{i}\prod_{j(\neq i)}p_{ij}^{a_{ij}}(1-p_{ij})^{1-a_{ij}}
\end{equation} 

\noindent having defined $p_{ij}\equiv\frac{x_{i}y_{j}}{1+x_{i}y_{j}}$, after having posed $x_i\equiv e^{-\alpha_i}$, $y_i\equiv e^{-\beta_i}$. The likelihood function is

\begin{equation}
\ln\mathcal{L}_{DCM}=\sum_{i}\left(k_i^{out}\ln x_i+k_i^{in}\ln y_i\right)-\sum_{i=1}^N\sum_{j(\neq i)=1}^N\ln(1+x_iy_j)
\end{equation}

\noindent and the maximum of the likelihood prescription becomes

\begin{eqnarray}
\left\{ \begin{array}{l}
k_i^{out}=\langle k_i^{out}\rangle^*=\sum_{j(\neq i)}\frac{x_i^*y_j^*}{1+x_i^*y_j^*},\:\forall\:i,\nonumber\\
k_i^{in}=\langle k_i^{in}\rangle^*=\sum_{j(\neq i)}\frac{x_j^*y_i^*}{1+x_j^*y_i^*},\:\forall\:i.
       \end{array} \right.
\end{eqnarray}

Once the unknown variables are numerically determined, the expected value of any adjacency matrix entry becomes 

\begin{equation} 
\langle a_{ij}\rangle^*=p_{ij}^*=\frac{x_{i}^*y_{j}^*}{1+x_{i}^*y_{j}^*}
\end{equation} 

\noindent and can be used to calculate the expected value of any other topological quantity of interest.

\subsection{Directed Configuration Model with Distances (DDCM)}

This model is one of the two main novelties that we introduce in the present paper. The DDCM Hamiltonian consists of the DCM Hamiltonian with one constraint more: a global quantity taking into account the information carried by the geographic distances, i.e.

\begin{eqnarray}
H(A,\:\vec{\theta})&=&\sum_{i=1}^N(\alpha_{i}k_i^{out}+\beta_ik_i^{in})+\gamma\sum_{i=1}^N\sum_{j(\neq i)=1}^Na_{ij}d_{ij}=\nonumber\\
&=&\sum_{i=1}^N\sum_{j(\neq i)=1}^N(\alpha_i+\beta_j+\gamma d_{ij})a_{ij}.
\end{eqnarray}

The information about the geographical distances is condensed in a global index, fixing the total sum of the connected vertices' distances. The reason why we introduce this model is that, according to recent results \cite{IEEEhowto:myfilling}, the spatial effects measured by the quantity $W_{tot}\equiv\sum_{i}\sum_{j(\neq i)}a_{ij}d_{ij}=\sum_{i}\sum_{j(>i)}(a_{ij}+a_{ji})d_{ij}$, even if weak, are not reproduced by the DCM. The DDCM reproduces those effects by construction, and we wil later introduce a way to quantify the corresponding information gain. Note that the Hamiltonian is again linear in the adjacency matrix entries: so, the probability of a given configuration factorizes again as the product $P(A|\vec{\theta})=\prod_{i}\prod_{j(\neq i)}p_{ij}^{a_{ij}}(1-p_{ij})^{1-a_{ij}}$, but with a different $p_{ij}$ coefficient:

\begin{eqnarray}
\langle a_{ij}\rangle=p_{ij}\equiv\frac{x_{i}y_{j}z^{d_{ij}}}{1+x_{i}y_{j}z^{d_{ij}}}
\label{pijddcm}
\end{eqnarray}

\noindent (and where the Lagrange multipliers have been reabsorbed into the hidden variables defintion: $x_i\equiv e^{-\alpha_i},\:y_i\equiv e^{-\beta_i},\:\forall\:i,\:z\equiv e^{-\gamma}$). The likelihood function is

\begin{eqnarray}
\ln\mathcal{L}_{DDCM}&=&\sum_{i}\left(k_i^{out}\ln x_i+k_i^{in}\ln y_i\right)+W_{tot}\ln z+\nonumber\\
&-&\sum_{i=1}^N\sum_{j(\neq i)=1}^N\ln(1+x_iy_jz^{d_{ij}})
\end{eqnarray}

\noindent and the maximization of the likelihood function leads to the following system to be solved

\begin{eqnarray}
\left\{ \begin{array}{l}
k_i^{out}=\langle k_i^{out}\rangle^*=\sum_{j(\neq i)}\frac{x_i^*y_j^*(z^*)^{d_{ij}}}{1+x_i^*y_j^*(z^*)^{d_{ij}}},\:\forall\:i,\\
k_i^{in}=\langle k_i^{in}\rangle^*=\sum_{j(\neq i)}\frac{x_j^*y_i^*(z^*)^{d_{ij}}}{1+x_j^*y_i^*(z^*)^{d_{ij}}},\:\forall\:i,\\
W_{tot}=\langle W_{tot}\rangle^*=\sum_{i}\sum_{j(\neq i)}\frac{d_{ij} x_i^*y_j^*(z^*)^{d_{ij}}}{1+x_i^*y_j^*(z^*)^{d_{ij}}},
       \end{array} \right.
\end{eqnarray}

\noindent where every index runs from $1$ to $N$. Once the previous system has been solved, we can use the $p_{ij}^*$ in (\ref{pijddcm}) to evaluate the expected value of all the topological quantities of interest. Note that, by posing $\gamma\equiv0$ ($z\equiv1$), we recover the usual DCM.

\subsection{Reciprocal Configuration Model (RCM)}

The third null model we consider is the Reciprocal Configuration Model (RCM) \cite{IEEEhowto:mymethod}. This model was defined to take into account the topological information encoded into the reciprocity structure of the observed network \cite{IEEEhowto:mygrandcanonical,IEEEhowto:myreciprocity}. It was shown that the RCM succeeds in reproducing almost perfectly all the triadic motifs of the WTW \cite{IEEEhowto:mymotifs}, which are instead not reproduced by the DCM. The RCM will therefore compete with the DDCM in improving the fit to the real network. The Hamiltonian of the RCM is the following

\begin{eqnarray}
H(A,\:\vec{\theta})=\sum_{i=1}^N(\alpha_i k_i^\rightarrow+\beta_i k_i^\leftarrow+\gamma_i k_i^\leftrightarrow)
\end{eqnarray}

\noindent and, unlike the previous two ones, it is not linear in the adjacency matrix entries. In fact, the imposed constraints are the three degree sequences, respectively defined as: the {\it non-reciprocated out-degree sequence}, where $k_{i}^{\rightarrow}=\sum_{j(\neq i)}a_{ij}^{\rightarrow}\equiv\sum_{j(\neq i)}a_{ij}(1-a_{ji})$, the {\it non-reciprocated in-degree sequence}, where $k_{i}^{\leftarrow}=\sum_{j(\neq i)}a_{ij}^{\leftarrow}\equiv\sum_{j(\neq i)}a_{ji}(1-a_{ij})$, the {\it reciprocated degree sequence}, where $k_{i}^{\leftrightarrow}=\sum_{j(\neq i)}a_{ij}^{\leftrightarrow}\equiv\sum_{j(\neq i)}a_{ij}a_{ji}$. All the above sequences are defined in terms of non-linear combinations of the $a_{ij}$s \cite{IEEEhowto:mymethod,IEEEhowto:mygrandcanonical,IEEEhowto:myreciprocity,IEEEhowto:mymotifs}. Nevertheless, the model is analitically solvable, the likelihood function is

\begin{eqnarray}
\ln\mathcal{L}_{RCM}&=&\sum_{i}\left(k_i^{\rightarrow}\ln x_i+k_i^{\leftarrow}\ln y_i+k_i^{\leftrightarrow}\ln z_i\right)+\nonumber\\
&-&\sum_{i=1}^N\sum_{j(\neq i)=1}^N\ln(1+x_iy_j+x_jy_i+z_iz_j)
\end{eqnarray}

\noindent and the maximization of the likelihood function leads to the following system to be solved:

\begin{eqnarray}
\left\{ \begin{array}{l}
k_i^{\rightarrow}=\langle k_i^{\rightarrow}\rangle^*=\sum_{j(\neq i)}\frac{x_i^*y_j^*}{1+x_i^*y_j^*+x_j^*y_i^*+z_i^*z_j^*},\:\forall\:i,\\
k_i^{\leftarrow}=\langle k_i^{\leftarrow}\rangle^*=\sum_{j(\neq i)}\frac{x_j^*y_i^*}{1+x_i^*y_j^*+x_j^*y_i^*+z_i^*z_j^*},\:\forall\:i,\\
k_i^{\leftrightarrow}=\langle k_i^{\leftrightarrow}\rangle^*=\sum_{j(\neq i)}\frac{z_i^*z_j^*}{1+x_i^*y_j^*+x_j^*y_i^*+z_i^*z_j^*}.\:\forall\:i.
       \end{array} \right.
\label{rcmsys}
\end{eqnarray}

\noindent where, now, we have three different probability coefficients: $(p_{ij}^{\rightarrow})^*$, $(p_{ij}^{\leftarrow})^*$, $(p_{ij}^{\leftrightarrow})^*$, respectively the generic addendum of the three equations above.

\subsection{Global Reciprocity Model (GRM)}

This model is a simplified version of the RCM \cite{IEEEhowto:symmetry}, where the reciprocity structure of the network is condensed in a general quantity, i.e. the total number of reciprocated links: $L^{\leftrightarrow}=\sum_{i}k_{i}^{\leftrightarrow}=\sum_{i}\sum_{j(\neq i)}a_{ij}a_{ji}$. It can be obtained from the RCM by posing $\gamma_i\equiv \alpha_i+\beta_i+\delta$. The constraints become: $k_i^{out}=\langle k_i^{out}\rangle^*$, $k_i^{in}=\langle k_i^{in}\rangle^*,\:\forall\:i$ and $L^{\leftrightarrow}=\langle L^{\leftrightarrow}\rangle^*$. The likelihood function is

\begin{equation}
\ln\mathcal{L}_{GRM}=\sum_{i}\left(k_i^{out}\ln x_i+k_i^{in}\ln y_i\right)+L^{\leftrightarrow}\ln z^2-\ln Z
\end{equation}

\noindent with $\ln Z=\sum_{i=1}^N\sum_{j(\neq i)=1}^N\ln(1+x_iy_j+x_jy_i+x_ix_jy_iy_jz^2)$ and the maximization of the likelihood function leads to the following system to be solved:

\begin{eqnarray}
\left\{ \begin{array}{l}
k_i^{out}=\sum_{j(\neq i)}\frac{x_i^*y_j^*+x_i^*x_j^*y_i^*y_j^*(z^*)^2}{1+x_i^*y_j^*+x_j^*y_i^*+x_i^*x_j^*y_i^*y_j^*(z^*)^2},\:\forall\:i,\\
k_i^{in}=\sum_{j(\neq i)}\frac{x_j^*y_i^*+x_i^*x_j^*y_i^*y_j^*(z^*)^2}{1+x_i^*y_j^*+x_j^*y_i^*+x_i^*x_j^*y_i^*y_j^*(z^*)^2},\:\forall\:i,\\
L^{\leftrightarrow}=\sum_{i}\sum_{j(\neq i)}\frac{x_i^*x_j^*y_i^*y_j^*(z^*)^2}{1+x_i^*y_j^*+x_j^*y_i^*+x_i^*x_j^*y_i^*y_j^*(z^*)^2}.
       \end{array} \right.
\end{eqnarray}

The GRM can be also considered as an enriched version of the DCM, to which the information about the global reciprocity has been added.

\section{Statistical validation}

We have a list of four null models to calculate the expected value of the topological quantities of interest. The second main novelty we introduce in this paper is the identification of a statistically correct procedure to compare these null models and to choose the most effective, among them.

\subsection{Likelihood ratio}

A first attempt could be that of comparing the likelihood values in their stationary points: the higher the value, the better the model to describe the considered network. A more quantitative way of testing the effectivness of two competing null models (say $NM_i$ and $NM_j$) is the calculation of their likelihood ratio, simply defined as

\begin{equation}
LR_{NM_i/NM_j}\equiv\frac{\ln\mathcal{L}_{NM_i}(\vec{\theta}_i)}{\ln \mathcal{L}_{NM_j}(\vec{\theta}_j)}
\end{equation}

\noindent where the symbols $\vec{\theta}_i$ and $\vec{\theta}_j$ indicate the two different sets of Lagrange multipliers. However, the likelihood ratio test suffers from three severe limitations \cite{IEEEhowto:likeratio}.

The first one lies in the fact that the null models have to be nested: the $i$-hypotesis has to be a special case of the $j$-hypotesis. Even if the DCM and the DDCM are nested and also the GRM and the DCM are nested as well, the RCM and the DDCM are not nested.

The second reason lies in the number of parameters. In fact, as the number of parameters rises, the agreement between the model and the observed network increases, too. So, even considering nested models, we could arbitrarily improve the $i$-hypotesis by simply adding more and more constraints. The drawback of this procedure is the risk of overfitting. 

The third reason lies in the number of models that can be tested: only two alternative hypoteses can be compared. We could only compare the effectiveness of two models at a time, ignoring the others and not carrying out a global comparison to choose from the whole set of models.

So, we need a criterion to choose among more than two competing null models, possibly not nested, and which discounts the number of parameters used to define them.

\subsection{Akaike Information Criterion (AIC)}

Indeed, the \emph{Akaike Information Criterion} suits our needs for selecting among several models \cite{IEEEhowto:aka,IEEEhowto:model_selection,IEEEhowto:model_selection2} by simply prescribing to calculate the following quantity

\begin{equation}
AIC^*_{NM_i}\equiv2K_{NM_i}-2\ln\mathcal{L}(\vec{\theta}^*)_{NM_i}
\end{equation}

\noindent i.e. the difference between (twice) the number of parameters of the null model $i$, $NM_i$, and (twice) its log-likelihood, evaluated in its maximum for every considered null model. For the four considered cases, we have: $K_{DCM}=2N$, $K_{DDCM}=2N+1$, $K_{GRM}=2N+1$, $K_{RCM}=3N$. Then, the recipe prescribes to choose the null model with the lowest AIC. 

\subsection{Akaike weights}

AIC simply tell us which model is the best, among those considered in the set. However, to quantify the improvement in choosing the best model with respect to the others, the so called \emph{Akaike weights} can also be computed, defined as

\begin{equation}
w_{NM_i}^{AIC}\equiv\frac{e^{-\frac{\Delta_{NM_i}}{2}}}{\sum_{r=1}^R e^{-\frac{\Delta_{NM_r}}{2}}}
\end{equation}

\noindent where $\Delta_{NM_i}\equiv AIC_{NM_i}^*-\min\{AIC_r^*\}_{r=1}^R$, being $R$ the total number of considered null models. The models with substantial support should have $\Delta\leq 2$, the models with less support should have $4\leq\Delta\leq7$ and models with $\Delta>10$ have essentially no support \cite{IEEEhowto:model_selection,IEEEhowto:model_selection2}.

The Akaike weights can be interpreted as the probability that the considered model is, in fact, the best one. Confidence intervals can also be built, reducing the number of models which could be considered as valid candidates \cite{IEEEhowto:model_selection,IEEEhowto:model_selection2,IEEEhowto:model_selection3,IEEEhowto:model_selection4}.

\subsection{Bayesian Information Criterion (BIC)}

Exactly as for the AIC, another quantity can be calculated and used to define the weights of the considered models: the \emph{Bayesian Information Criterion}. The only difference lies in the term to be discounted from the maximized likelihood:

\begin{equation}
BIC^*_{NM_i}\equiv K_{NM_i}\ln n-2\ln\mathcal{L}(\vec{\theta}^*)_{NM_i};
\end{equation}

\noindent the first addendum accounts not only for the number of parameters, $K$, but also for the cardinality of the sample, $n$. In our case, $n=N(N-1)$, because we are considering directed matrices. The \emph{BIC weights} are defined analogously:

\begin{equation}
w_{NM_i}^{BIC}\equiv\frac{e^{-\frac{\Delta_{NM_i}^{B}}{2}}}{\sum_{r=1}^R e^{-\frac{\Delta_{NM_r}^{B}}{2}}}
\end{equation}

\noindent where $\Delta_{NM_i}^{B}\equiv BIC_{NM_i}^*-\min\{BIC_r^*\}_{r=1}^R$, being $R$ the total number of considered null models. Criteria to interpret the BIC weights similar to those stated above hold. It is commonly said the AIC favours the model with the highest number of parameters and that BIC, on the other hand, could be more restrictive, favouring one of the models with less parameters \cite{IEEEhowto:model_selection,IEEEhowto:model_selection2}. Since the discussion is still topical, we have presented and compared both.

\section{Degree-degree correlations, reciprocity, filling}

In order to integrate our analysis with previous results in the literature, we will also study the performance of the various models in reproducing some specific structural properties. We consider the \emph{in-degree correlations} and the \emph{out-degree correlations} defined as

\begin{equation}
k_{i}^{in/in}=\frac{\sum_{j(\neq i)}a_{ji}k_j^{in}}{k_i^{in}},\:k_{i}^{out/out}=\frac{\sum_{j(\neq i)}a_{ij}k_j^{out}}{k_i^{out}},
\end{equation}

\noindent the \emph{reciprocity} $r$ \cite{IEEEhowto:reciprocity} and the \emph{filling} $f$ \cite{IEEEhowto:myfilling}:

\begin{equation}
r=\frac{L^{\leftrightarrow}}{L},\:f=\frac{W_{tot}-min}{max-min}
\end{equation}

\noindent where 

\begin{equation}
max\equiv\sum_{i=1}^Ld_{i}^{\downarrow}\:\:\mbox{and}\:\:d^{\downarrow}\equiv(d_1,\:d_2\dots d_{N(N-1)})
\end{equation}

\noindent with the pairs of distances ordered in decreasing order, $d_1=d_2\geq d_3=d_4\geq\dots\geq d_{N(N-1)}$, and

\begin{equation}
min\equiv\sum_{i=1}^Ld_{i}^{\uparrow}\:\:\mbox{and}\:\:d^{\uparrow}\equiv(d_{N(N-1)}\dots\:d_2,\:d_1)
\end{equation}

\noindent with the pairs of distances ordered in increasing order. The filling was recently introduced to measure the tendency of a network to fill the euclidean space where it is embedded \cite{IEEEhowto:myfilling}. This goal is accomplished by measuring how the geographic distances are distributed over the topological links.

Different methods were chosen to compare the effectiveness of the four null models in explaining the three quantities above. For the degree-degree correlations, the scatter plots of the observed and the expected $k_i^{in/in}$ and $k_i^{out/out}$ are shown. 

For the reciprocity and the filling a different quantity was defined, to incorporate in a single index the observed and the expected values under the chosen null model ($NM$) \cite{IEEEhowto:myreciprocity,IEEEhowto:myfilling}:

\begin{equation}
\rho_{NM}=\frac{r-\langle r\rangle_{NM}}{1-\langle r\rangle_{NM}},\:\phi_{NM}=\frac{f-\langle f\rangle_{NM}}{1-\langle f\rangle_{NM}},
\end{equation}

\noindent where, e.g.

\begin{equation}
\langle r\rangle_{NM}=\frac{\sum_i\sum_{j(\neq i)} p_{ij}^{NM}p_{ji}^{NM}}{\sum_{j(\neq i)}\sum_i p_{ij}^{NM}}
\end{equation}

\noindent and $p_{ij}^{NM}$ indicates that the choice of a particular null model $NM$ is nothing more than the choice of the corresponding probability coefficients for the adjacency matrix entries. Note also that both $\rho$ and $\phi$ are normalized between 1 and $-1$.

\section{Results and Discussion}

\begin{figure}[t!]
\centering
\hspace{1mm}\includegraphics[scale=0.55]{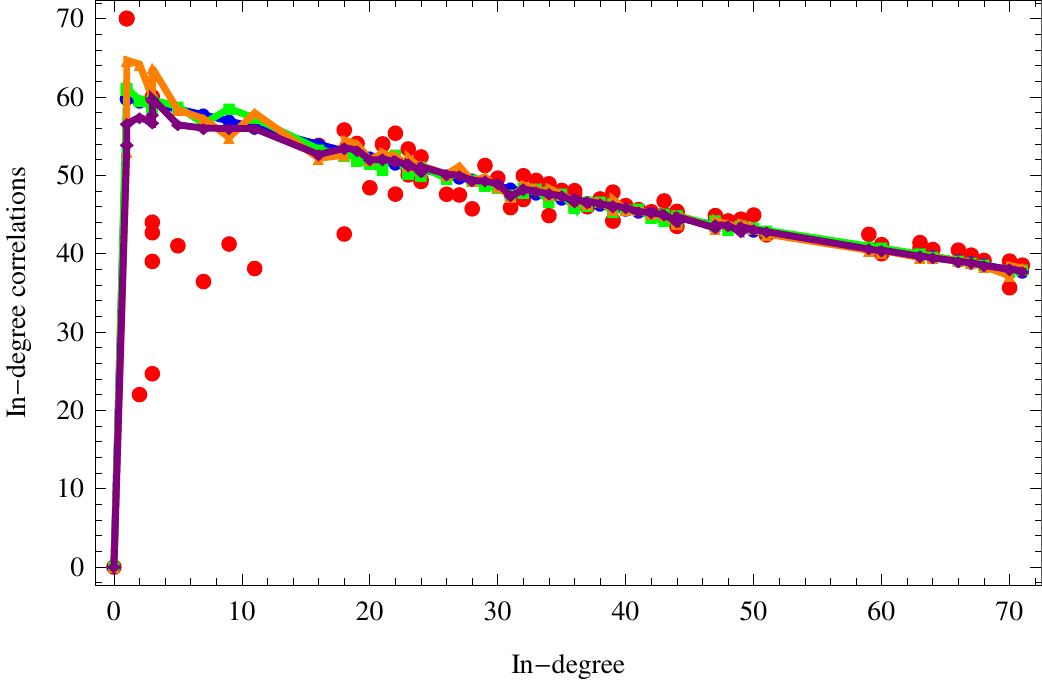}\\
\hspace{2mm}\includegraphics[scale=0.55]{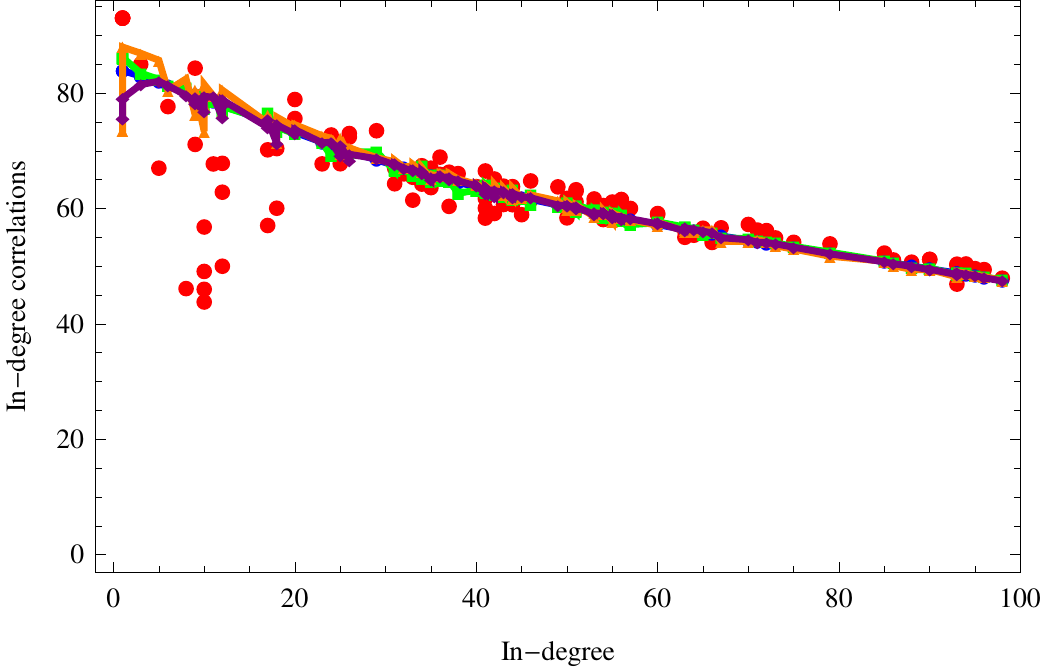}\\
\includegraphics[scale=0.55]{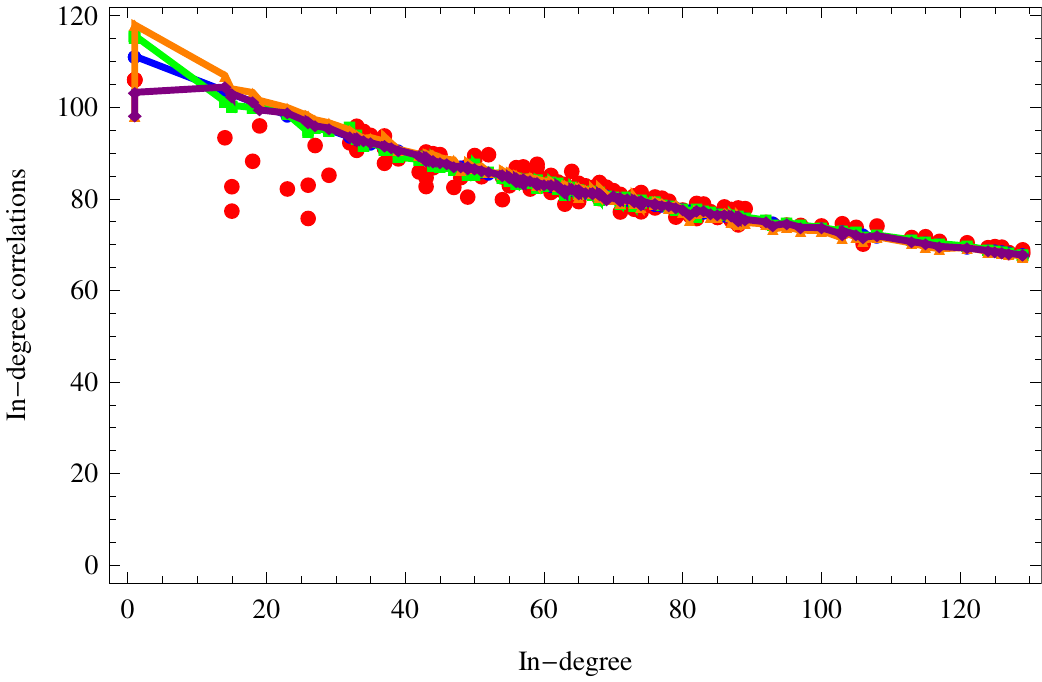}\\
\includegraphics[scale=0.55]{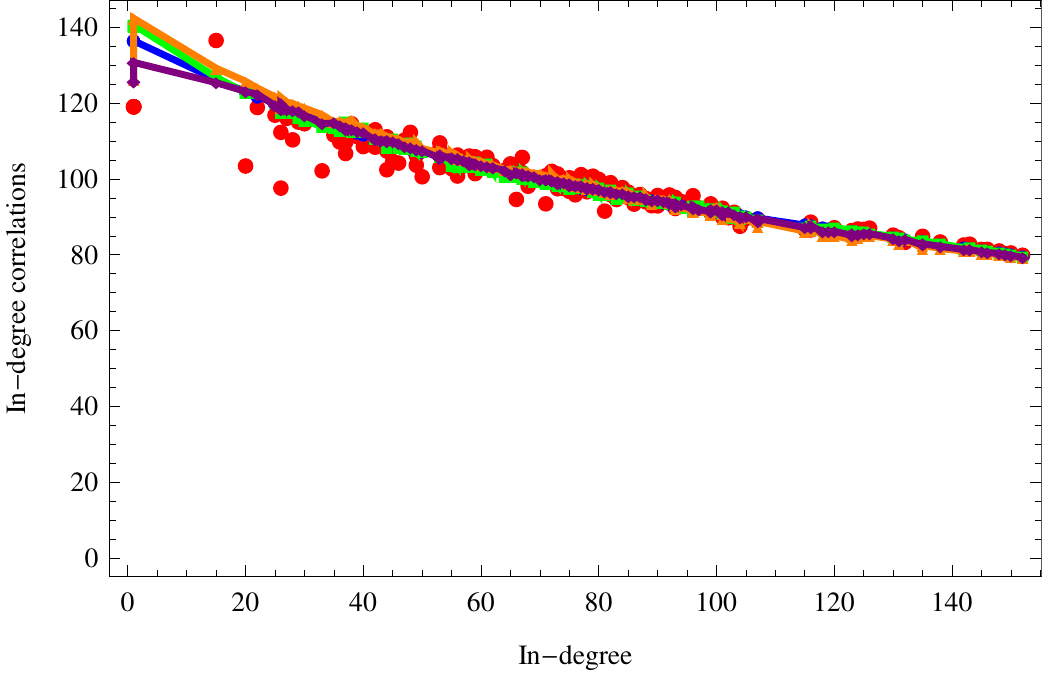}\\
\includegraphics[scale=0.55]{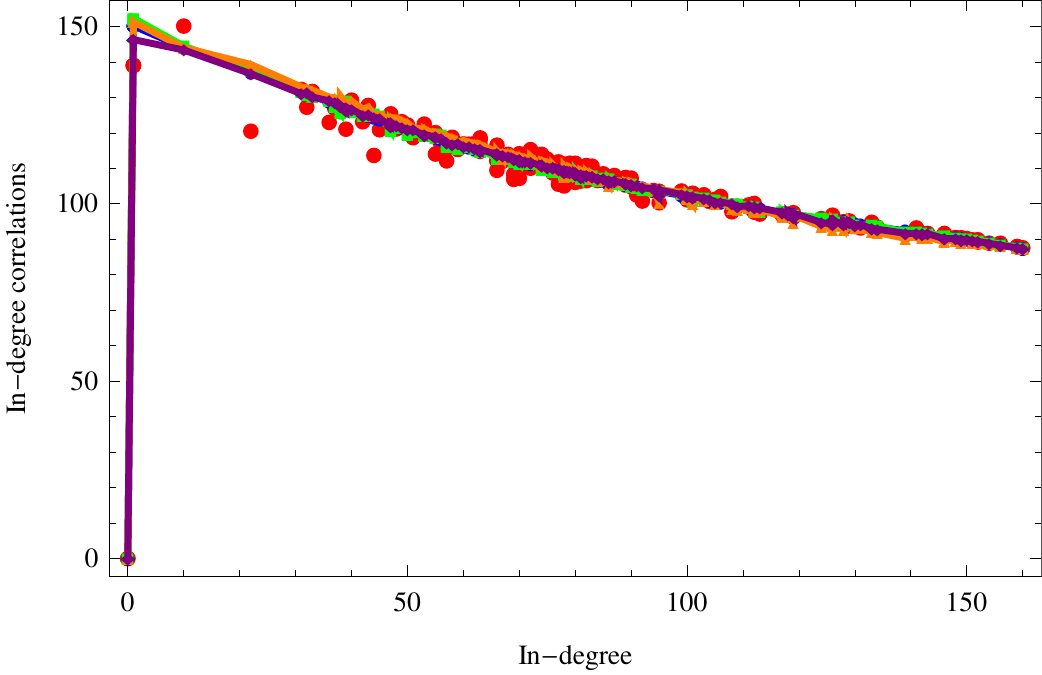}
\caption{Observed (red) and expected (colored) $k_i^{in/in}$ vs $k_i^{in}$ for the five decades 1950, 1960, 1970, 1980, 1990. The colored trends represent: the DCM (blue), the DDCM (green), the RCM (orange) and the GRM (violet).}
\label{scatters}
\end{figure}

\subsection{Degree-degree correlations}

The degree-degree correlations are analysed by comparing the observed values of $k_i^{in/in}$ and $k_i^{out/out}$ with their expected values under the four, considered null models. The results are shown in fig. \ref{scatters}. As a general consideration, the four null models qualitatively perform in approximately the same way, as the superposition of the four colored trends shows. 

However, by carefully looking at the quantitative differences we find out that the expected trends under the DCM are improved by all the remaining three models which preserve the DCM statistics but also add one, or more, contraints. 

In fact, they become less smooth, by following the irregularities of the observed, scattered points. In this respect, the RCM seems to perform best, by showing the larger deviations from the DCM trend: on the other side, the DDCM and the GRM, by both adding one parameter to the DCM constraints, are very similar to each other.

From a temporal perspective, the 1950 and the 1960 (first and second row in fig. \ref{scatters}) are the sparsest network with the most scattered trends, showing the least agreement between the observations and the expectations.

\begin{figure}[t!]
\centering
\hspace{1mm}\includegraphics[scale=0.55]{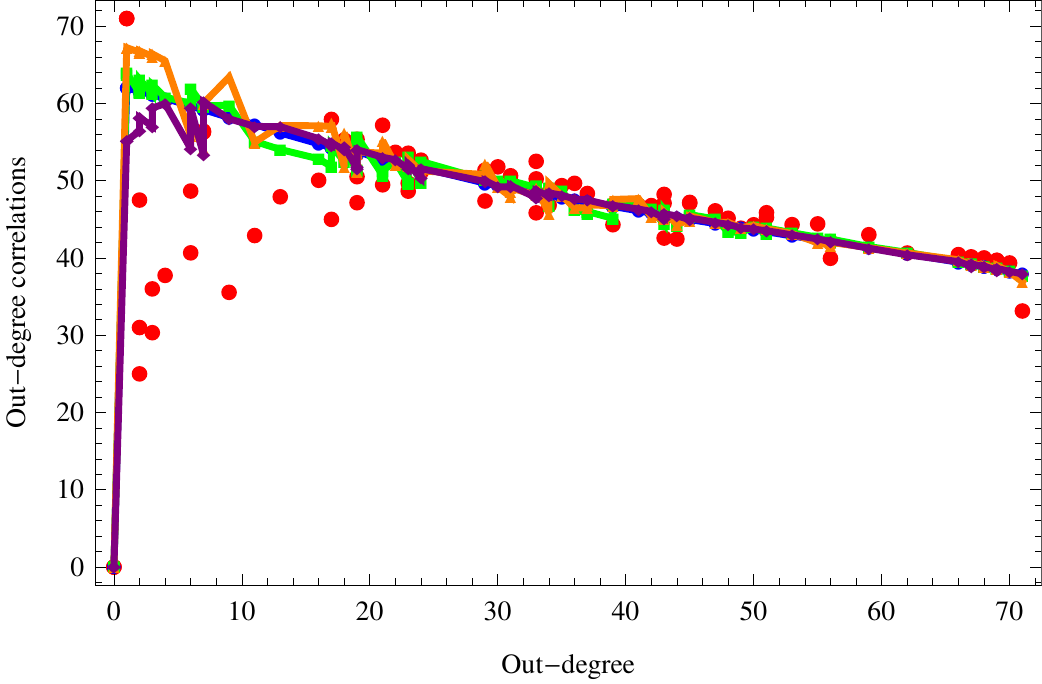}\\
\includegraphics[scale=0.55]{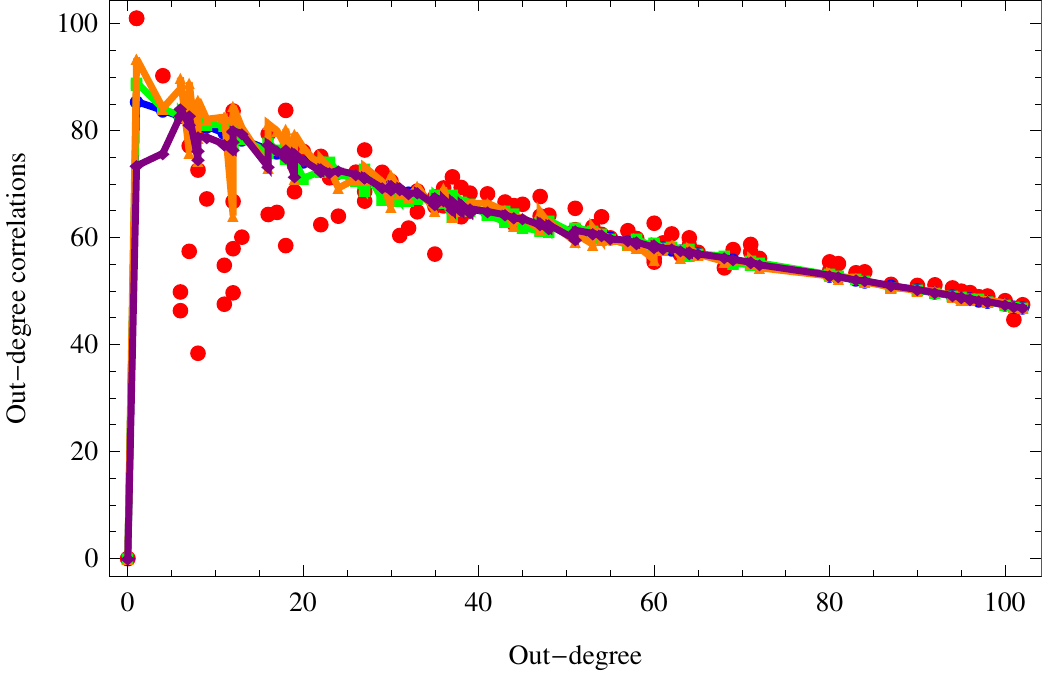}\\
\includegraphics[scale=0.55]{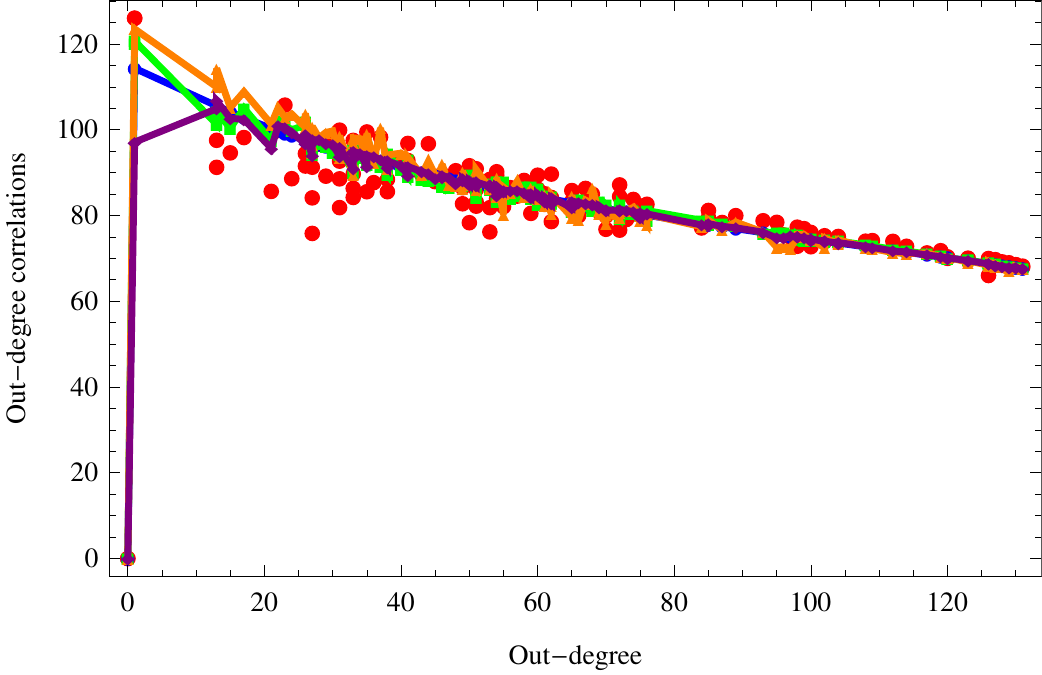}\\
\includegraphics[scale=0.55]{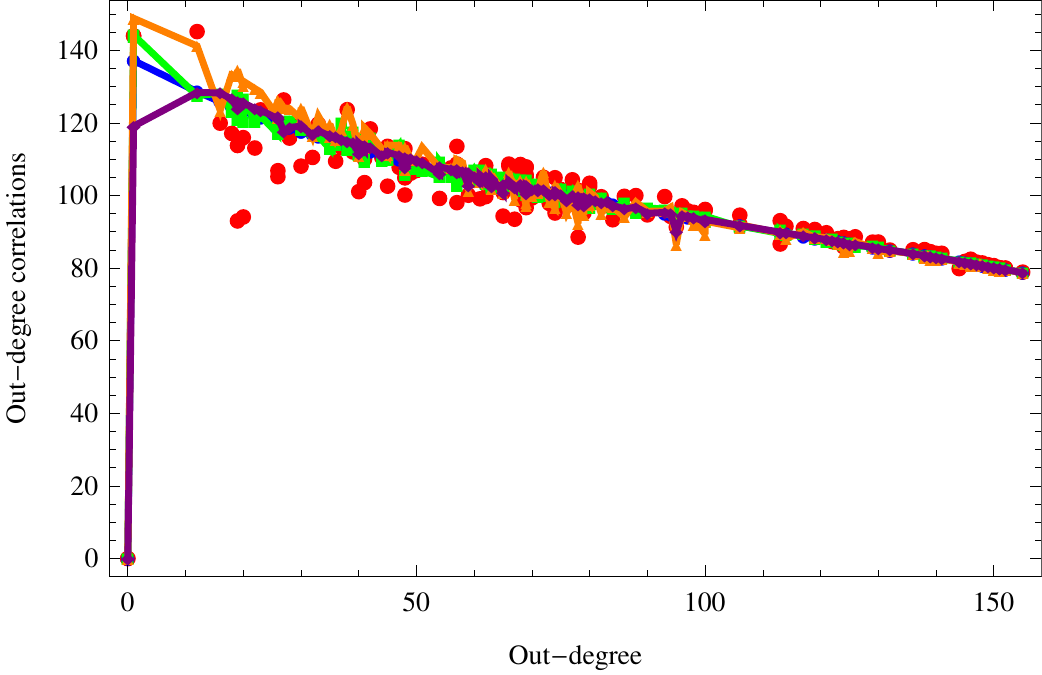}\\
\includegraphics[scale=0.55]{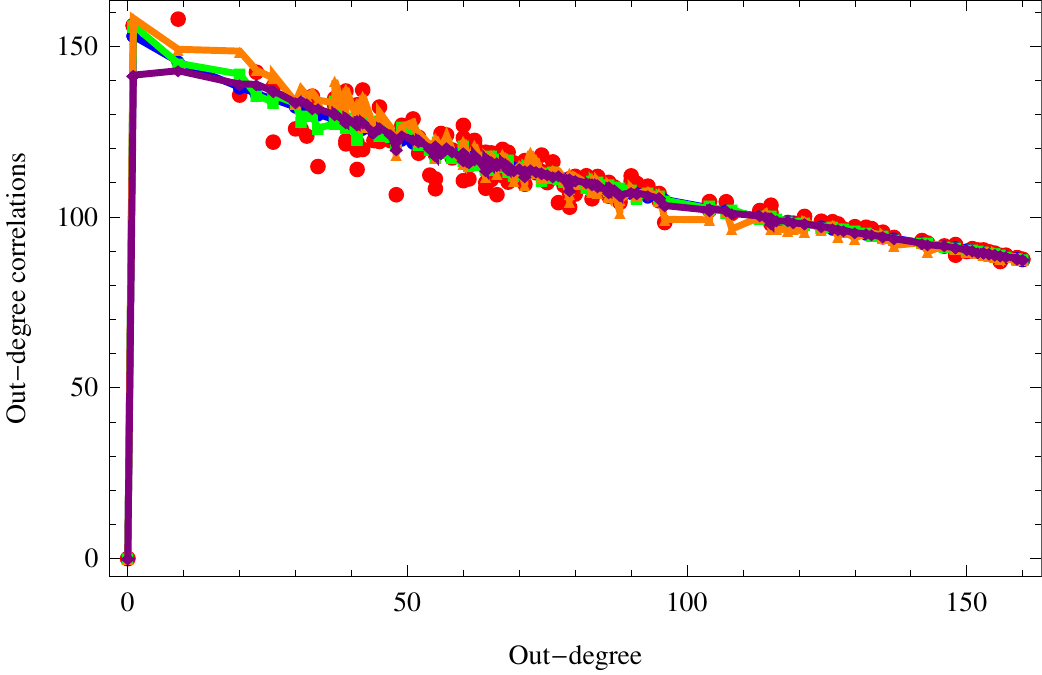}
\caption{Observed (red) and expected (colored) $k_i^{out/out}$ vs $k_i^{out}$ for the five decades 1950, 1960, 1970, 1980, 1990. The colored trends represent: the DCM (blue), the DDCM (green), the RCM (orange) and the GRM (violet).}
\label{scatters}
\end{figure}

\subsection{Reciprocity and filling}

By the definition of $\rho$ and $\phi$, it is clear that both $|\rho|\leq1$ and $|\phi|\leq1$. In fact, the denominator is the normalization constant not contributing to the sign of the quantity itself. So the sign of $\rho$ and $\phi$ is decided only by the relative magnitude between the observed values, $r$ and $f$, and their expectation: a positive sign indicates a stronger than expected tendency to reciprocate or to fill the embedding space. On the other hand, a negative sign indicates a weaker than expected tendency to reciprocate or to fill the embedding space.

\begin{figure}[t!]
\centering
\hspace{3mm}\includegraphics[width=2.5in]{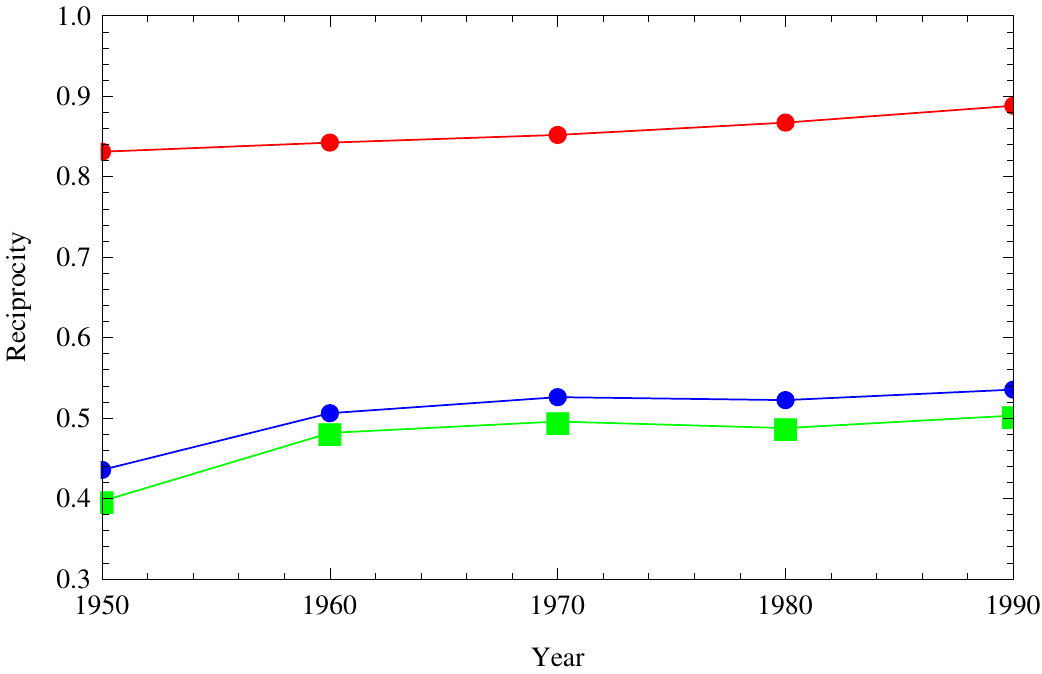}
\caption{Observed reciprocity, $r$ (red), $\rho_{DCM}$ (blue) and $\rho_{DDCM}$ (green). By definition, $\rho_{RCM}=\rho_{GRM}=0$.}
\label{fig:rec}
\end{figure}

Let us consider the reciprocity. From fig. \ref{fig:rec} and table \ref{tab:RecV} is clear that the DCM and the DDCM perform almost the same in exaplining the observed reciprocity: the positive sign of $\rho$ indicates the tendency of the network to reciprocate more than expected (both under the DCM and the DDCM) but the addition of the information about the geographic distances improves the agreement between the observed $r$ and the expected $r$, signalled by a lower value of the corresponding $\rho$. Note that both the RCM and the GRM incorporate the information encoded into the reciprocity: so, by definition, $\rho_{RCM}=\rho_{GRM}=0$.

Now, let us consider the filling. In this case, the DCM, the GRM and the RCM perform the same, indicating the tendency of the network to fill the embedding space less than expected (in fact, the sign is negative). However, in this case the addition of the infomation about the (global or local) reciprocity structure does not seem to add anything more to what predicted by the directed degree sequences alone. In this case, the model incorporating the information carried by the filling is the DDCM, for which $\phi_{DDCM}=0$ by definition. 

\subsection{AIC and BIC criteria}

The previous subsections have shown semi-quantitative attempts to test the effectiveness of the four null models in explaining the observed patterns. The above results appear to qualitatively confirm that the strongest factor shaping the WTW topology is the degree sequence \cite{IEEEhowto:pre1}, but small improvements can be made either by adding spatial factors \cite{IEEEhowto:myfilling} or reciprocity effects \cite{IEEEhowto:mymotifs}. In order to quantify how ``small'' the improvements are, and whether are statistically significant, we use the criteria introduced in sections III B. and III D. Table \ref{tab:Bic} shows the AIC and the BIC values for the null models considered so far. Apart from 1950, AIC favours the RCM for all the years, the model with the highest number of parameters which specifies the local reciprocity structure of the observed network. This is compatible both with the trends showed in the scatter plots and with the value of $\rho_{RCM}$ which, being zero, provides the best prediction for the expected value of $r$. For the filling prediction, the RCM and the GRM perform the same.

\begin{table}[t!]
\centering
\caption{$r$ and $\rho$ for the four null models.}
\begin{tabular}{|c|c|c|c|c|c|c|}
\hline
$\mbox{\bf Year}$ & $ \mathbf{r} $ & $ \mathbf{\rho _{DCM}} $ & $\mathbf{\rho_{DDCM}} $ & $ \mathbf{\rho _{RCM}} $ & $ \mathbf{\rho _{GRM}} $\\
\hline
\hline
$\mbox{1950}$ & $0.83 $ & $0.43$ & $0.39$ & 0 & 0\\
\hline
$\mbox{1960}$ & $0.84 $&$0.50 $ &$0.48$ & 0 & 0\\
\hline
$\mbox{1970}$ & $0.85$&$0.53 $ & $0.50$ & 0 & 0\\
\hline
$\mbox{1980}$ & $0.86$&$0.52 $ &$0.49$ & 0 & 0\\
\hline
$\mbox{1990}$ & $0.89$&$0.54 $ & $0.50$ & 0 & 0\\
\hline
\end{tabular}
\label{tab:RecV}
\end{table}

BIC, on the other side, always favours the GRM, the model which adds to the DCM only the global information about the reciprocity. This is compatible both with the value of $\rho_{GRM}$ and with the filling prediction: we already commented the small difference in the predicted values of the degree-degree correlations under the RCM and the GRM.

So, by looking only at the AIC and BIC values we are left with two possible models to choose between: RCM and GRM. Let us calculate the weights, as shown in table \ref{tab:BicW}. Apart from 1950, AIC weights always favour the RCM, which is the model with the highest probability to be the most correct. For the year 1950, the RCM and the GRM compete and should be both retained \cite{IEEEhowto:model_selection,IEEEhowto:model_selection2}. On the other side, BIC weights always favour the GRM which seems to be accurate enough to give the best prediction.

With respect to the DCM, the DDCM (i.e. the DCM with the addition of the geographic distances) is actually better, as signalled by a lower value of AIC and BIC. Since the degree sequences are known to be positively correlated to the world countries GDPs, this means that by considering the distances in addition to the GDPs improves the prediction of the model. 

\begin{figure}[t!]
\centering
\includegraphics[width=2.5in]{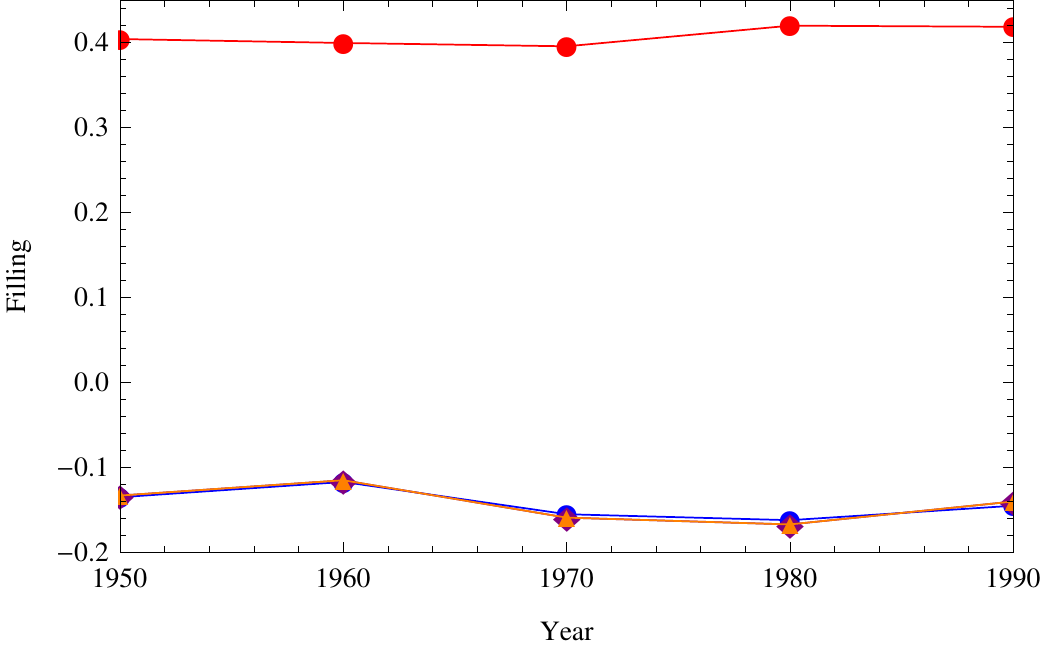}
\caption{Observed filling, $f$ (red), $\phi_{DCM}$ (blue), $\phi_{RCM}$ (orange), $\phi_{GRM}$ (purple). By definition, $\rho_{DDCM}=0$.}
\label{fig:fill}
\end{figure}

\begin{table}[t!]
\centering
\caption{$f$ and $\phi$ for the four null models.}
\begin{tabular}{|c|c|c|c|c|c|}
\hline
$\mbox{\bf Year}$ & $ \mathbf{f} $ & $ \mathbf{\phi_{DCM}} $ & $\mathbf{\phi_{DDCM}}$ & $ \mathbf{\phi_{RCM}}$ & $ \mathbf{\phi_{GRM}}$\\
\hline
\hline
$\mbox{1950}$ & $0.40$ & $-0.13$ & 0 & $-0.13$ & $-0.13$\\
\hline
$\mbox{1960}$ & $0.40$ & $-0.11$ & 0 & $-0.11$ & $-0.11$\\
\hline
$\mbox{1970}$ & $0.39$ & $-0.15$ & 0 & $-0.15$ & $-0.15$\\
\hline
$\mbox{1980}$ & $0.42 $ & $-0.16$ & 0 & $-0.16$ & $-0.16$\\
\hline
$\mbox{1990}$ & $0.41$ &$ -0.14$ & 0 & $-0.14$ & $-0.14$\\
\hline
\end{tabular}
\label{tab:fill}
\end{table}

What about the GRM? As the DDCM, also the GRM adds only one parameter to the DCM: in fact, they have the same number of parameters, i.e. $2N+1$. However, in the DDCM it is sufficient to introduce the parameter $z$ to consider the whole matrix of geographic distances that, in turn, affect every single probability connection $p_{ij}$. On the contrary, in the GRM the parameter $z$ only introduces one quantity, $L^{\leftrightarrow}$, and remains the same for every pair of nodes. On the basis of this apparent convenience, we could be tempted to choose the DDCM as the best between the two, but this is not supported by the two criteria which confirm the GRM as the best model between them (and, eventually, among all). This remains valid also for the year 1950, even if BIC indicates DDCM as a preferable model with respect to RCM: anyway, GRM outperforms both. This indicates that in order to have a good prediction of the WTW in 1950, the whole, local reciprocity structure is redundant: the global information about the distances could be a \emph{better} choice. But the \emph{best} choice is represented by the global information about the reciprocity structure.

So, given the DCM constraints (the in-degree and out-degree sequences), the next best choice to impose an additive constraint does not involve the distances between countries but the global reciprocity structure of the trade-exchanges network. In other words, given the GDPs of the world countries, a better choice than the common one would be the definition of a gravity model incorporating the information about the reciprocal trade-exchanges.

\begin{table}[t!]
\centering
\caption{AIC and BIC for the considered null models (rounded to the nearest integer).}
\begin{tabular}{|c|c|c|c|c|}
\hline
$\mbox{\bf Year}$ & $ \mathbf{AIC_{DCM}} $ & $ \mathbf{AIC_{DDCM}} $ & $ \mathbf{AIC_{RCM}}$ & $ \mathbf{AIC_{GRM}}$\\
\hline
\hline
$\mbox{1950}$ & $ \simeq5172$ & $\simeq4796$ &$\simeq4646$ & $\simeq 4645$\\
\hline
$\mbox{1960}$ & $\simeq9840$ &$\simeq9360$ & $\simeq8576$ & $\simeq 8593$\\
\hline
$\mbox{1970}$ & $\simeq16816$ & $\simeq15818$& $\simeq14218$ & $\simeq 14406$\\
\hline
$\mbox{1980}$ & $\simeq20539$ & $\simeq19135$ &$\simeq17435$ & $\simeq 17680$\\
\hline
$\mbox{1990}$ & $\simeq20496$ &$\simeq19170$ & $\simeq17165$ & $\simeq 17492$\\
\hline
\hline
$\mbox{\bf Year}$ & $ \mathbf{BIC_{DCM}} $ & $ \mathbf{BIC_{DDCM}} $ & $\mathbf{BIC_{RCM}} $ & $ \mathbf{BIC_{GRM}}$ \\
\hline
\hline
$\mbox{1950}$ & $\simeq 5594$ & $\simeq 5221$ & $\simeq5280$ & $\simeq 5070$ \\
\hline
$\mbox{1960}$ & $\simeq 10471$ &$\simeq 9994$ & $\simeq9523$ & $\simeq 9227$ \\
\hline
$\mbox{1970}$ & $\simeq 17640$ & $\simeq 16645$ &$\simeq15454$ & $\simeq 15233$\\
\hline
$\mbox{1980}$ &$\simeq21523$ & $\simeq 20122$ & $\simeq18911$ & $\simeq 18667 $\\
\hline
$\mbox{1990}$ & $\simeq 21537$ &$\simeq 20215$ & $\simeq18727$  & $\simeq 18537 $\\
\hline
\end{tabular}
\label{tab:Bic}
\end{table}

\section{Conclusion}

In this paper we have considered four null models to analyse five decades of the World Trade Web, represented as binary, directed networks. The adopted approach was different from that of the gravity-models (or the zero-inflated gravity models, created to manage binary networks), making use of the exponential random graph formalism. We have therefore rephrased the problem of distances by suitably defining structural constraints in term of the adjacency matrix elements.

Starting from the DCM Hamiltonian we have considered more, and different, topological quantities to test the effectiveness of the geographic distances in explaining the binary structure of the WTW and to compare them with the other types of chosen constraints.

The geographic distances were introduced by means of a global index and added to the DCM, but we found only a slight improvement of the latter. In the same way, another global index was introduced to consider the global reciprocity structure of the network. What emerges from the statistical criteria used to indentify the best model is that, actually, two models compete and (unless considering the multimodel averaging inference alternative \cite{IEEEhowto:model_selection,IEEEhowto:model_selection2}) should both be retained: the RCM and the GRM (the first one already performed successfully in the motifs analysis of the WTW \cite{IEEEhowto:mymotifs}). In the same way, if we calculate the AIC and the BIC weights between the DDCM and the GRM (having the same number of parameters but exploiting different information: all the distances or only the total number of reciprocal links), the latter always performs better than the former (with probability 1).

It should be noted that, in principle, geography effects might already be present in the degree sequence, so that controlling for the latter automatically controls (at least partially) for the distances. Therefore, the correct way to interpret our results should be that the role of geography, if present, is almost entirely encoded within the degree sequence, so the additional explicit inclusion of distance constraint does not improve the modeling significantly. However, we do not expect distances to be significantly encoded into the (reciprocated or not) degree sequence, for various reasons. First of all, the degrees of countries are known to be depend strongly on the GDP \cite{IEEEhowto:myinterplay}. The latter varies over many orders of magnitude, while distances vary only within a narrow range. Secondly, degrees are local (vertex-specific) properties, whereas distances are pairwise (edge-specific) properties. By preserving only the degrees, the DCM breaks the possible original associations between connectivity and distances, but still reproduces the WTW well. Finally, as clear from (\ref{pijddcm}), the inclusion of distances in the DDCM does not introduce a probabilistic dependence between a link and its reciprocal one, a dependence which is instead produced by the inclusion of reciprocity in the RCM and GRM. Therefore, we do not expect distance effects to be encoded in the reciprocated degree sequences as well, because the strong reciprocity of the latter could not be explained by the DDCM, and not even by gravity models.

Our result conclusively show that although spatial effects are indeed present in the WTW topology they are entirely dominated by the non-spatial effects determined by the reciprocity. This suggests to prefer the reciprocity structure of the network (local, with the RCM, or global, with the GRM) to the geographic distances, in order to obtain more precise  predictions about the trade-exchanges. This, in turn, implies that the information coming from the GDPs, in a gravity model framework, should be sustained by some other economic indicator about the reciprocal trade activity of the involved countries.

\begin{table}[t!]
\centering
\caption{AIC weights and BIC weights for the considered null models.}
\begin{tabular}{|c|c|c|c|c|}
\hline
$\mbox{\bf Year}$ & $ \mathbf{w_{DCM}^{AIC}} $ & $ \mathbf{w_{DDCM}^{AIC}} $ & $\mathbf{w_{RCM}^{AIC}} $ &$ \mathbf{w_{GRM}^{AIC}}$ \\
\hline
\hline
$\mbox{1950}$ & $0$ & $0$ &  $ 0.36 $ & $0.64$\\
\hline
$\mbox{1960}$ & $0$ &$0$ & $ 1 $ & $ 0$\\
\hline
$\mbox{1970}$ & $0$ & $0$& $ 1 $ &$0$\\
\hline
$\mbox{1980}$ & $0$ & $0$ & $ 1 $ & $ 0$\\
\hline
$\mbox{1990}$ & $0$ &$0$ & $ 1 $ & $ 0 $ \\
\hline
\hline
$\mbox{\bf Year}$ & $ \mathbf{w_{DCM}^{BIC}} $ & $ \mathbf{w_{DDCM}^{BIC}} $ & $\mathbf{w_{RCM}^{BIC}} $ & $ \mathbf{w_{GRM}^{BIC}}$ \\
\hline
\hline
$\mbox{1950}$ & $0$ & $0$ & $ 0$ & $1$\\
\hline
$\mbox{1960}$ & $0$ &$0$ & $ 0$ & $1$\\
\hline
$\mbox{1970}$ & $0$ & $0$& $ 0$ & $1$\\
\hline
$\mbox{1980}$ & $0$ & $0$ & $ 0$ & $1$\\
\hline
$\mbox{1990}$ & $0$ &$0$ & $ 0$ & $1$\\
\hline
\end{tabular}
\label{tab:BicW}
\end{table}

\section*{Acknowledgment}

D. G. acknowledges support from the Dutch Econophysics Foundation (Stichting Econophysics, Leiden, the Netherlands) with funds from beneficiaries of Duyfken Trading Knowledge BV, Amsterdam, the Netherlands.

T. S. acknowledges support from an ERC Advanced Investigator Grant.




\begin{thebibliography}{1}

\bibitem{IEEEhowto:giorgio}
G.~Fagiolo, ``The International-Trade Network: Gravity Equations and Topological Properties'', J. Econ. Interact. Coord., vol. 5, pp. 1-25 (2010).

\bibitem{IEEEhowto:gravity}
G.-J.~M. Linders, M.~J. Burger and F.~G. van Oort, ``A rather empty world: the many faces of distance and the persistent resistance to international trade'', Cam. J. Reg. Econ. Soc., vol. 1, pp. 439-458 (2008).

\bibitem{IEEEhowto:giorgio2}
M.~Duenas and G.~Fagiolo, ``Modeling the International-Trade Network: A Gravity Approach'', arXiv:1112.2867 (2012).

\bibitem{IEEEhowto:mywegrec}
T.~Squartini, F.~Picciolo, F.~Ruzzenenti and D. Galraschelli, ``Reciprocity of weighted networks'', arXiv:1208.4208 (2012)

\bibitem{IEEEhowto:mygrandcanonical}
D.~Garlaschelli and M.~I. Loffredo, ``Multi-species grandcanonical models for networks with reciprocity '', Phys. Rev. E, vol. 73, 015101(R) (2006).

\bibitem{IEEEhowto:myreciprocity}
D.~Garlaschelli and M.~I. Loffredo, ``Patterns of link reciprocity in directed networks'', Phys. Rev. Lett., vol. 93, 268701 (2004).

\bibitem{IEEEhowto:mymotifs}
T.~Squartini and D.~Garlaschelli, ``Triadic motifs and dyadic self-organization in the World Trade Network'', Lec. Not. Comp. Sci., vol. 7166, pp. 24-35 (2012).

\bibitem{IEEEhowto:myinterplay}
D.~Garlaschelli and M.~I. Loffredo, ``Interplay between topology and dynamics in the World Trade Web '', Eur. Phys. J. B, vol. 57, pp. 159-164 (2007).

\bibitem{IEEEhowto:mylikelihood}
D.~Garlaschelli and M.~I. Loffredo, ``Maximum likelihood: extracting unbiased information from complex networks'', Phys. Rev. E, vol. 78, 015101(R) (2008).

\bibitem{IEEEhowto:gledi}
K.~S. Gleditsch, ``Expanded Trade and GDP Data'', Jour. Conf. Res., vol. 46, 712 (2002).

\bibitem{IEEEhowto:myfilling}
F.~Ruzzenenti, F.~Picciolo, R.~Basosi and D.~Garlaschelli, ``Space filling in the World Trade Web: measures and null models '', arXiv:1207.1791 (2012).

\bibitem{IEEEhowto:mymethod}
T.~Squartini and D.~Garlaschelli, ``Analytical maximum-likelihood method to detect patterns in real networks'', New. J. Phys., vol. 13, 083001 (2011).

\bibitem{IEEEhowto:HL}
P.~Holland and S.~Leinhardt, ``Sociological Methodology'', \relax San Francisco, USA: Heise Ed. (1975).

\bibitem{IEEEhowto:WF}
S.~Wasserman and K.~Faust, ``Social Network Analysis'', \relax New York, USA: Cambridge University Press (1994).

\bibitem{IEEEhowto:newman_expo}
J.~Park and M.~E.~J. Newman, ``The statistical mechanics of networks'', Phys. Rev. E, vol. 70, 066117 (2004).

\bibitem{IEEEhowto:shannon}
C.~Shannon, ``A Mathematical Theory of Communication'', Bell System Tech. Jour., vol. 27, pp. 379-423 (1948).

\bibitem{IEEEhowto:jaynes}
E.~T. Jaynes, ``Information Theory and Statistical Mechanics'', Phys. Rev., vol. 106, pp. 620-630 (1957).

\bibitem{IEEEhowto:pre1}
T.~Squartini, G.~Fagiolo and D.~Garlaschelli, ``Randomizing world trade. I. A binary network analysis '', Phys. Rev. E, vol. 84, 046117 (2011).

\bibitem{IEEEhowto:symmetry}
F.~Ruzzenenti, D.~Garlaschelli and R.~Basosi, ``Complex Networks and Symmetry II: Reciprocity and Evolution of World Trade'', Symmetry, vol. 2, pp. 1710-1744 (2010).

\bibitem{IEEEhowto:likeratio}
D.~R. Cox and D.~V. Hinkley, ``Theoretical Statistics'', \relax New York, USA: Chapman and Hall (1974).

\bibitem{IEEEhowto:aka}
H.~Akaike, ``A new look at the statistical model identification'', IEEE Trans. Aut. Cont., vol. 19, pp. 716-723 (1974).

\bibitem{IEEEhowto:model_selection}
K.~P. Burnham and D.~R. Anderson, ``Model Selection and Inference: A Practical Information-Theoretical Approach'', \relax New York, USA: Springer-Verlag (1998).

\bibitem{IEEEhowto:model_selection2}
K.~P. Burnham and D.~R. Anderson, ``Multimodel inference: understanding AIC and BIC in Model Selection'', Soc. Met. Res., vol. 33, pp. 261-304 (2004).

\bibitem{IEEEhowto:model_selection3}
E.~J. Wagenmakers and S.~Farrell, ``AIC model selection using Akaike weights'', Psych. Bull Rev., vol. 11, pp. 192-196 (2004).

\bibitem{IEEEhowto:model_selection4}
J.~B. Johnson and K.~S. Omland, ``Model selection in ecology and
evolution'', Trends Ecol. Evol., vol. 9, pp. 101-108 (2004).

\bibitem{IEEEhowto:reciprocity}
M.~E.~J. Newman, S.~Forrest and J.~Balthrop, ``Email networks and the spread of computer viruses'', Phys. Rev. E, vol. 66, 035101(R) (2002).

\end{thebibliography}
%

\end{document}